# Switchable Photovoltaic Effect in Ferroelectric CsPbBr$_3$ Nanocrystals

Anashmita Ghosh[1], Susmita Paul[1], Mrinmay Das[1], Piyush Kanti Sarkar[1], Pooja Bhardwaj[2], Goutam Sheet[2], Surajit Das[1], Anuja Datta[1] and Somobrata Acharya[1]*

[1] School of Applied & Interdisciplinary Sciences, India Association for the Cultivation of Science, Jadavpur, Kolkata-700032, India

[2] Department of Physical Sciences, Indian Institute of Science Education and Research (IISER) Mohali, Sector 81, S. A. S. Nagar, Manauli P.O. 140306, India

Corresponding Author: camsa2@iacs.res.in

**Abstract:** Ferroelectric all-inorganic halide perovskites nanocrystals with both spontaneous polarizations and visible light absorption are promising candidates for designing functional ferroelectric photovoltaic devices. Three-dimensional halide perovskite nanocrystals have the potential of being ferroelectric, yet it remains a challenge to realize ferroelectric photovoltaic devices which can be operated in absence of an external electric field. Here we report that a popular all-inorganic halide perovskite nanocrystal, CsPbBr$_3$, exhibits ferroelectricity driven photovoltaic effect under visible light in absence of an external electric field. The ferroelectricity in CsPbBr$_3$ nanocrystals originates from the stereochemical activity in Pb (II) lone pair that promotes the distortion of PbBr$_6$ octahedra. Furthermore, application of an external electric field allows the photovoltaic effect to be enhanced and the spontaneous polarization to be switched with the direction of the electric field. Robust fatigue performance, flexibility and prolonged photoresponse under continuous illumination are potentially realized in the zero-bias conditions. These finding establishes all-inorganic halide perovskites nanocrystals as potential candidates for designing novel photoferroelectric devices by coupling optical functionalities and ferroelectric responses.

## 1. Introduction

Ferroelectricity is a remarkable phenomenon that plays a significant role in modern technologies. Ferroelectrics undergo a transition from a high-symmetry structure to a low-symmetry state with a spontaneous electric polarization below a transition temperature. Such spontaneous polarization occurs below the Curie temperature and can be switched using an external electric field.[1,2] Ferroelectric oxide perovskites[3] have shown potential applications in ferroelectric random access memories,[4] switchable ferroelectric diodes,[5,6] electromechanical transducers,[7–9] non-volatile memories,[10,11] actuators,[12] and in the fields of high-*K* materials.[13] Ferroelectricity driven photovoltaic effect is an emerging area of research, which has been primarily observed only in traditional oxide ferroelectric materials.[14–17] Obvious advantages in using ferroelectric photovoltaic materials include the possibility of generating a large open circuit voltage, facile separation of charge carriers, slow recombination, and long carrier diffusion lengths.[18] Recently, ferroelectricity induced photovoltaic effects have been observed in hybrid halide perovskites, in particular MAPbI$_3$.[18–20] In the context of hybrid halide perovskites, ferroelectricity originates owing to two obvious reasons. The polar MA (CH$_3$NH$_3^+$) cation organic units carry dipoles, which is the prerequisite of any ferroelectric material. In addition, Pb(II) containing perovskites are known to give rise to ferroelectricity arising from its lone pair, which helps to distort it from a centrosymmetric position, thereby generating a dipole locally.[21–24] Even very small distortions from centrosymmetric to non-centrosymmetric structures can potentially lead to an induced dipole in the perovskite crystal lattice. Thus, a necessary condition for such a perovskite material to be ferroelectric is non-centrosymmetric and polar, so that it can exhibit spontaneous polarization. Resultant spontaneous polarization can create a built-in electric field to separate the photogenerated carriers spatially without applying external bias.[25,26] As a result, steady short-circuit current and large open-circuit voltage can be obtained in the orientation of spontaneous polarization. Such ferroelectricity induced photovoltaics has emerged as an indispensable branch of new optoelectronic devices.

To date, the ferroelectric photovoltaic effect has been observed in oxide perovskites such as bismuth ferrite (BiFeO$_3$), lead zirconate titanate (Pb(ZrTi)O$_3$) and barium titanate (BaTiO$_3$) showing the polarization-sensitive activity with a promise for polarized light detection.[27–29] A unique characteristic of ferroelectric photovoltaic devices is that the photocurrent direction can be reversed by changing the spontaneous polarization direction of a ferroelectric material with the application of an electric field. Recent developments into ferroelectric photovoltaic include above band gap photovoltaics of BiFeO$_3$ (BFO) using symmetric platinum top electrodes,[15] bulk photovoltaic effect (BPVE) of BiFeO$_3$,[28] non-volatile memory using BiFeO$_3$.[30] Beyond these oxide ferroelectrics, giant switchable photovoltaic effect was observed from hybrid halide perovskite MAPbI$_3$ under illumination.[31,32] Anomalous photovoltaic effect was observed from MAPbBr$_3$(or MAPbI$_3$ and CsPbBr$_3$) thin films, which was not caused by ferroelectricity, rather formation of tunneling junctions as a result of ion migration in the randomly dispersed polycrystalline film resulted in accumulation of photovoltage.[33] Biaxial 2D hybrid perovskitesferroelectric,(iso-pentylammonium)$_2$(ethylammonium)$_2$Pb$_3$I$_{10}$ (PEPI), exhibited self-driven photodetection in the visible range owing to the large spontaneous polarization.[34] Self-powered UV photodetection was demonstrated using single crystal of 2D wide-band gap hybrid perovskite ferroelectric (BPA)$_2$PbBr$_4$[35] and 2D ferroelectric layered lead chloride hybrid perovskite, EA$_4$Pb$_3$Cl$_{10}$.[36] These self-powered devices rely on the use of asymmetric electrodes that leads to the separation of electrons and holes or poling by means of an external electric field to induce ferroelectric domain switching. Recently, all-inorganic perovskite materials have drawn a great attention due to their potential application in photovoltaics and optoelectronics.[37-39] However, it remains a challenge to realize ferroelectric photovoltaics with symmetric electrodes without poling by an external electric field and to the best of our knowledge, there is not yet any report of ferroelectric photovoltaics using all-inorganic halide perovskite nanocrystals (NCs) that operates in zero-bias condition.

Here ferroelectric photovoltaics is realized in a popular all-inorganic CsPbBr$_3$ NCs using symmetric electrodes in absence of an external electric field. We show that the CsPbBr$_3$ NCs exhibit ferroelectricity with a notable saturated polarization of ∼0.15 μC/cm$^2$ and high Curie temperature of 465 K. The built-in electric field induced by ferroelectric polarization promotes the separation of photogenerated carriers, which provide an opportunity to fabricate ferroelectric photovoltaics that operates in zero bias conditions. The photovoltaic performance of the CsPbBr$_3$ NCs can be further improved by poling with an external electric field. Our studies of the poling direction dependency reveal polarization reversal with the direction of the external electric field. Flexibility, robust fatigue performance and prolonged photoresponse under continuous illumination were potentially realized. These results stimulate the exploration of all-inorganic halide perovskite NCs to be a class of promising candidate for the novel ferroelectric photovoltaic applications.

## 2. Results and Discussion

We synthesized CsPbBr$_3$ NCs using lead bromide (PbBr$_2$) and cesiumoleate (Cs-OA) in presence of mixture of long chain capping ligands oleic acid (OA) and oleylamine (OLAm) at 165°C under N$_2$ atmosphere.[40] Transmission electron microscope (TEM) image shows cubic morphology of the CsPbBr$_3$ NCs with an average dimension of ~20 ± 5nm (Figure 1a). High resolution TEM (HRTEM) image of NCs reveals well-resolved lattice planes implying crystalline nature of the NCs (Figure S1). The HRTEM image shows an interplanar spacing of ~0.29 ± 0.05 nm corresponding to the (220) planes of bulk orthorhombic crystal structure (COD # 1533062). The selected area electron diffraction (SAED) pattern of the NCs shows diffraction spots corresponding to the orthorhombic crystal phase (Figure S2). Powder X−ray diffraction (XRD) pattern of CsPbBr$_3$ NCs also shows reflections corresponding to orthorhombic phase (COD #

1533062; space group (62): Pbnm; a = 8.2070 Å, b = 8.2550 Å, c = 11.7590 Å) (Figure S3). Energy dispersive X–ray spectroscopy (EDS) measurements in TEM reveals atomic ratio of Cs:Pb:Br ~ 1:1:3, which confirms the CsPbBr$_3$ chemical composition of the NCs (Figure S4).

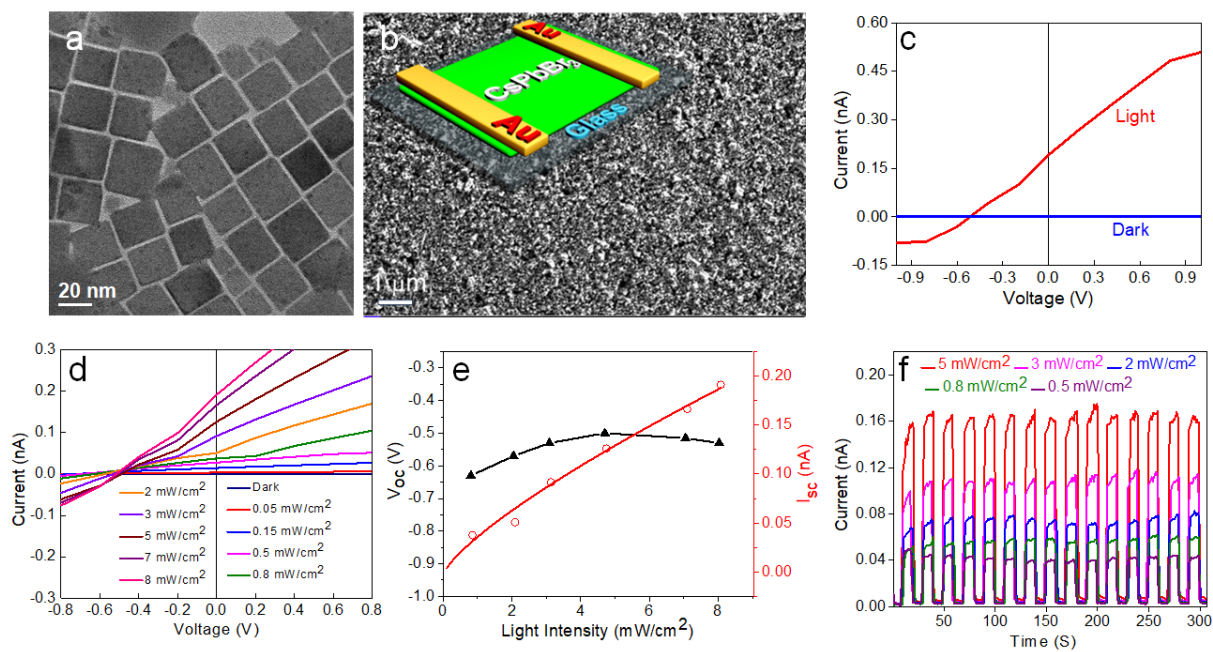

**Figure 1.** (a) TEM image of CsPbBr$_3$ NCs. (b) SEM image of the drop casted CsPbBr$_3$ NCs on glass substrate showing a compact layer of NCs. Symmetric electrode lateral device structure is shown in the inset. (c) Current-voltage curves measured in dark (blue) and under light (red) at zero-bias condition. (d) Current-voltage curves measured with different light intensities at zero-bias condition. (e) Variation of open circuit voltage (black) and short circuit current (red) with light intensity. (f) The time-dependent light on-off response with different light intensities at zero-bias condition.

As synthesized NCs were drop-casted on glass substrates to fabricate the photovoltaic devices. Scanning electron micrograph (SEM) image revealed a compact layer of NCs upon drop casting (Figure 1b). The device was fabricated using a lateral symmetric structure Au/CsPbBr$_3$/Au (Figure 1b, inset), where the Au electrodes were fabricated on top of compact layers of CsPbBr$_3$ NCs. The current (I) versus voltage (V) characteristics revealed photovoltaic behaviour at zero-bias under 8 mW/cm$^2$ illumination without any poling of the NCs layer by an external electric field (Figure 1c). An open circuit voltage ($V_{OC}$) of 0.51 V and short circuit current ($I_{SC}$) of 0.19 nA were obtained demonstrating a clear photovoltaic effect under light illumination. Devices fabricated with Au/Ti/CsPbBr$_3$/Ti/Au or Ag/ CsPbBr$_3$ on Polyethylene terephthalate (PET)/Ag also showed similar photovoltaic response suggesting non-contamination of the NCs layer upon electrode deposition (Figure S5, S6). Such photovoltaic effect has been found in inorganic oxide ferroelectrics, such as BiFeO$_3$[28] and BaTiO$_3$,[41] owing to the bulk photovoltaic effect (BPVE) due to their non-centrosymmetric nature which can create a built-in electric field owing to the spontaneous polarization to spatially separate the photogenerated carriers without applying external bias.[15,42,43] The observation of zero-bias photovoltaic effect with symmetric electrodes is important since previous reports on the halide perovskite solar cells did not show any photovoltaic effect with the symmetric lateral electrode structure containing the same work function or without poling by an external electric field.[33] Such a remarkable $V_{OC}$ and $I_{SC}$ obtained from the CsPbBr$_3$ NCs at the zero-bias condition is comparable to the previously reported hybrid halide perovskite ferroelectrics, such as (C$_2$H$_5$NH$_3$)$_2$(CH$_3$NH$_3$)$_2$Pb$_3$Br$_{10}$[44] and (3-pyrrolinium)-(CdCl$_3$),[45] or (BPA)$_2$PbBr$_4$[35] which were capable of converting photons into current in self-power mode without external bias.

We observe that the photovoltaic response of the device depends on the intensity of the light (Figure 1d). When the light intensity increases at zero-bias conditions, both $V_{OC}$ and $I_{SC}$ change initially (Figure 1e). Although the $V_{OC}$ saturates at 0.51 V, no saturation is observed for $I_{SC}$ up to light intensity of 8 mW/cm$^2$. Increasing light intensity generated more electron-hole pairs which are separated by the internal field leading to the increase of $I_{SC}$. This indicates that most of the photogenerated excitons dissociated to free charges in the CsPbBr$_3$ NCs layer and the charges were efficiently collected by the symmetric electrodes. The dependency of photocurrent with the light intensity can be fitted to a power law, $I \propto P^n$, where n= 0.77 determines the response of photocurrent to the light intensity.[46] The non-integer power law dependency of photocurrent on light intensity indicates an electron–hole generation and separation processes, as observed in other organic and inorganic photovoltaics devices.[47,48]

The CsPbBr$_3$ NCs can be potentially used to generate photocurrent considering its operation at zero-bias conditions. The time-dependent photoresponse on-off cycles under zero bias with variable light intensities are shown in Figure 1f. The photocurrent increases rapidly upon illumination indicating the spontaneous separation for photoinduced electrons and holes under zero-bias conditions. The time dependent photoresponse shows that the photocurrent rapidly reaches the stable state with a high value, and then drastically decreases to its initial value as the light is turned off. The photodetection is carried out for multiple light on-off cycles, which shows that the device can operate normally under zero-bias condition. It should be noted that the dark current is as low as ~pA. Thus, a large on-off ratio up to $10^3$ is evidenced, which is significant for achieving efficient photodetection at zero-bias condition. Such a photoresponse of CsPbBr$_3$ NCs under the symmetric junction barriers is unusual. The photodetection exhibits obvious light intensity dependent characters. The magnitude of photocurrent in the light on state increases with the light intensity and then reaches to the same low state when the light is turned off. The UV-vis absorption spectrum of CsPbBr$_3$ NCs reveals a band gap of 2.3 eV (Figure S7). The absorption edge is at around 535 nm, which resulted in from the near band-edge absorption of CsPbBr$_3$ NCs. Under white light illumination, CsPbBr$_3$ NCs are excited and the photovoltaic derives operate mainly from electron–hole pairs excited by incident light with energy higher than the band gap. Driven by the built-in electric field, the charge carriers are collected by the electrodes to generate the photocurrent, while the sub-bandgap states make little contribution to the photocurrent.

In order to probe that the zero-bias photovoltaics is intensely related to the ferroelectric origin, we first measured the temperature-dependent dielectric properties of the $CsPbBr_3$ NCs at a frequency ranging from 2 kHz to 1 MHz since the paraelectric-ferroelectric phase transition is accompanied by an obvious dielectric anomaly (Figure 2a). The ferroelectric nature of the transition is evident from the large dielectric anomalies at temperatures around the Curie temperature ($T_c$~465 K). The temperature-dependent real part ($\varepsilon_r$) of the dielectric constant at each frequency, including the higher frequencies, displays a pronounced dielectric peak around the $T_c$, which is associated with the phase transition in the $CsPbBr_3$ NCs. Similarly, $CsPbBr_3$ NCs at a frequency range from 2kHz to 1MHz exhibit a low dielectric loss (<0.1) at room temperature which increases gradually near $T_c$ (Figure S8). The $T_c$ ~ 465 K of $CsPbBr_3$ NCs is higher compared to the $CsPbBr_3$ QDs (298 K),[23] bulk $CsPbBr_3$ (403 K),[49] organic−inorganic ferroelectrics (400–440 K),[50–52] [TMAEA]$Pb_2Cl_6$ (412 K),[53] [$(CH_3)_3NCH_2I$]$PbI_3$ (312 K),[54] [4,4-difluorocyclohexylammonium]$_2PbI_4$ (377 K)[55] and inorganic oxide perovskite ferroelectric $BaTiO_3$ (393 K).[56] Such high $T_c$ makes the $CsPbBr_3$ NCs suitable for ferroelectric photovoltaic applications. The prominent dielectric properties of the $CsPbBr_3$ NCs conform to the Curie–Weiss law (Figure S9) and correspond to the anomalous behavior due to ferroelectric–paraelectric transition. According to the Curie–Weiss law, the reciprocal of $\varepsilon_r$ as a function of temperature is linear.[57] The intercept of the Curie–Weiss temperature obtained by extrapolation is estimated to be~ 465 K, which is similar to $T_c$. The ratio of Curie–Weiss constant ($C_{para}/C_{ferro}$) is calculated to be 0.66, which discloses the characteristics of second-order ferroelectric phase transition of the $CsPbBr_3$ NCs.

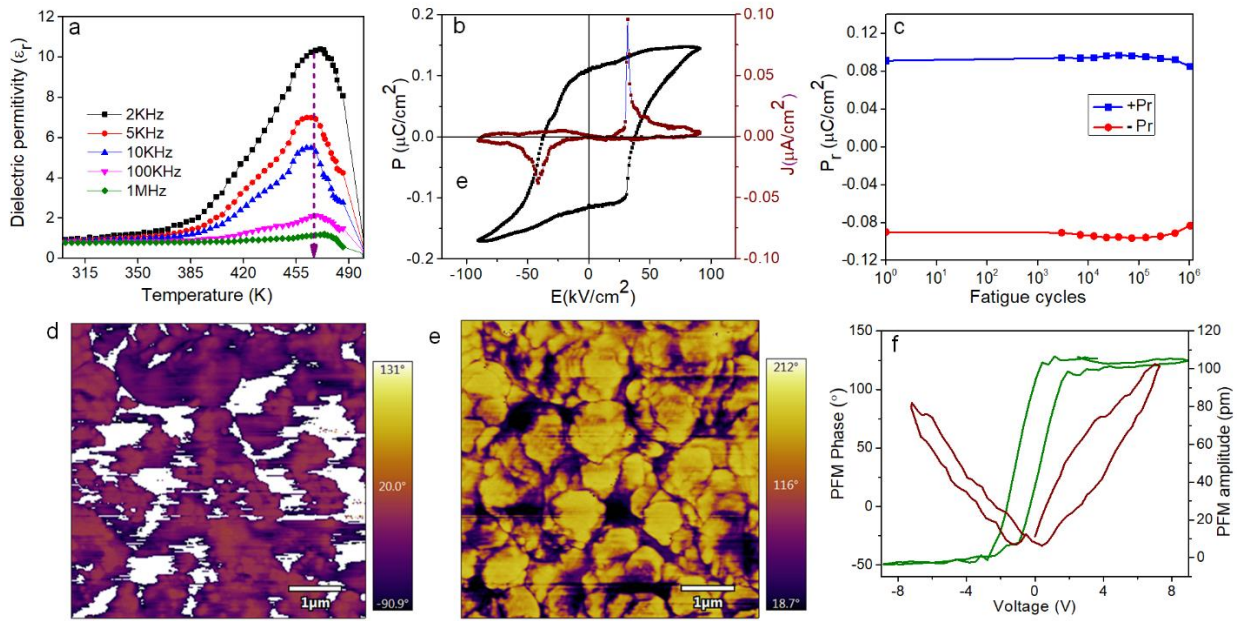

**Figure 2.** (a) Temperature dependence of the dielectric permittivity of $CsPbBr_3$ NCs at different frequencies obtained at room temperature. The inset shows the frequency range from 2KHz to 1MHz. (b) Ferroelectric P−E hysteresis loop (black) and corresponding J−E curve (wine) of $CsPbBr_3$ NCs. (c) Remanent polarization versus number of fatigue cycles for up and down polarization revealing no fatigue after $10^6$ cycles. (d) PFM phase image after the first electric poling with a tip bias of + 10 V. (e) PFM phase image after the second electric poling with a tip bias of − 10 V. (f) Hysteretic dependence of the lateral PFM phase (wine) and amplitude (green) with applied DC bias for $CsPbBr_3$ NCs.

To further examine the ferroelectricity in $CsPbBr_3$ NCs, polarization versus electric field (P-E) hysteresis loops were measured using the double wave method (Figure 2b).[23] The P-E loop shows well-developed, saturated hysteresis and signature of polarization switching confirming the ferroelectric nature of the $CsPbBr_3$NCs. The P-E hysteresis loop reveals a saturation polarization ($P_s$) of ~0.15 $\mu C/cm^2$, remanent polarization ($P_r$) of ~0.11$\mu C/cm^2$ and coercive field ($E_c$) of 37.5 kV/cm at room temperature. The degree of squareness ($P_s/P_r$) of the loop becomes nearly ≈ 1.36. Previously, Li et al. reported a maximum $P_s$ of 0.018 $\mu C/cm^2$ at temperature of 293 K in a pressed pelletized disc composed of $CsPbBr_3$ QDs. [23] To the best of our knowledge, such high $P_s$ value has been observed for the first time in $CsPbBr_3$ NCs at above the room temperature. The higher polarization of $CsPbBr_3$ NCs in comparison to the pelletized $CsPbBr_3$ QDs is attributed mainly to the orientation of the ordered $CsPbBr_3$ NCs which resulted in ferroelectric domain alignment as opposed to the randomly oriented domains in the pelletized CsPbBr$_3$ QDs. Additionally, the high dielectric breakdown threshold of the CsPbBr$_3$NCs allowed the application of higher fields than bulk pelletized disks of CsPbBr$_3$ QDs.[23]

To reveal the typical P-E loop resulted in from the polarization switching, we measured the switching current density versus electric field (J–E) loop for the CsPbBr$_3$ NCs (Figure 2b). Two switching current peaks around the E$_c$ are observed in the J–E curve. These two opposite peaks indicate two stable states with opposite polarization suggesting that the dipole moments can be switched by revering the external electric field.[58] The section of the J–E curve between the two current peaks resembles semiconductor nature, which is a characteristic of ferroelectric material. Low leakage switching current is also measured from the ferroelectric CsPbBr$_3$NCs indicating the efficacy of the thin-film capacitor device structures (Supporting Information Figure S10). Moreover, CsPbBr$_3$NCs exhibit excellent endurance against ferroelectric fatigue up to ~10$^6$ cycles (Figure 2c). The P$_r$ of CsPbBr$_3$ NCs as a function of the number of switching cycles show only <2 % deviation from the pristine value during the fatigue test.

The microscale domain structure is another essential property of polarity for the ferroelectrics, which has been probed by using the piezoresponse force microscopy (PFM).[59] The topography of the surface and corresponding amplitude image of the CsPbBr$_3$ NCs are shown in Figure S11. Domain imaging by out-of-plane phase contrast is evidenced in PFM phase image (Figure 2 d, e). We performed domain switching tests by applying a DC tip bias of ±10 V over 6 μm ×6 μm image area. A clear reversal of the~180° phase contrast is observed, demonstrating that the polarizations in CsPbBr$_3$NCs can be switched back and forth in the local scale (Figure 2 d, e). Along with the phase contrast imaging from PFM, local switching spectroscopy was performed to demonstrate the polarization switching behavior. Local PFM spectroscopic measurements show the butterfly loop and strong hysteresis behavior in amplitude and phase curves, respectively, providing additional evidence of ferroelectricity in CsPbBr$_3$NCs (Figure 2f).

Earlier, first-principles calculations demonstrated the ferroelectricity of all-inorganic CsPbF$_3$ perovskites driven by the lone pair of Pb (II).[24] Recently, ferroelectricity has been observed in all-inorganic perovskite CsGeI$_3$ nanostructures driven by stereochemical activity of lone pair in Ge (II).[59] Organic–inorganic hybrid halide perovskite DMAGeX$_3$ (X = Cl, Br, I) showed ferroelectricity owing to the stereochemically active $4s^2$ lone pair of Ge (II) and the ordering of organic cations.[60] Ferroelectricity in CsPbBr$_3$ QDs was also realized owing to the 6s lone pair electrons of Pb (II) that imparts distortion of the [PbBr$_6$] octahedra.[23] The CsPbBr$_3$QDs adopts a structural distortion with twisting and tilting of the [PbBr$_6$] octahedra during the cubic P$m\bar{3}m$ to orthorhombic P$na2_1$ phase transition. Hence, the observed ferroelectricity of CsPbBr$_3$ NCs could be assessed to the stereochemical activity of the lone pair of Pb (II) cations. Unlike the traditional semiconductor photovoltaic devices, where charge carriers separate at p−n junctions, ferroelectric domains in CsPbBr$_3$ NCs with its built-in electric field acts to separate the photo generated carriers to generate photovoltaic effect.

Further we measure the influence of the poling on the photovoltaic response of the CsPbBr$_3$ NCs. The halide perovskite devices showed an increased V$_{OC}$ and I$_{SC}$ after poling by a moderate electric field at room temperature (Figure 3a). Magnitude of V$_{OC}$ and I$_{SC}$ can be further increased by increasing the poling strength. Furthermore, the directions of V$_{OC}$ and I$_{SC}$ can be switched by reversing the direction of the electric field (Figure 3b). After negative poling by an electrical field of 7 V/μm for 5 min, V$_{OC}$ increased to −0.6 V whereas V$_{OC}$ showed switching to +0.21 V after positive poling (Figure 3c). Simultaneously, the I$_{SC}$ changes from 0.16 nA to −0.03 nA upon reversal of the poling direction. Figure 3c shows the photoresponse on-off cycles at zero bias and after poling with increasing electric fields in same direction. A gradual increase of the photocurrent is evidenced. Figure 3d represents the transient photoresponse of the device under zero-bias and after poling in opposite direction. An improved photoresponse can be obtained after poling which can be reversed by changing the direction of poling. The switchable nature of the photovoltaic effect indicates that it is related to the spontaneous polarization of CsPbBr$_3$ NCs.

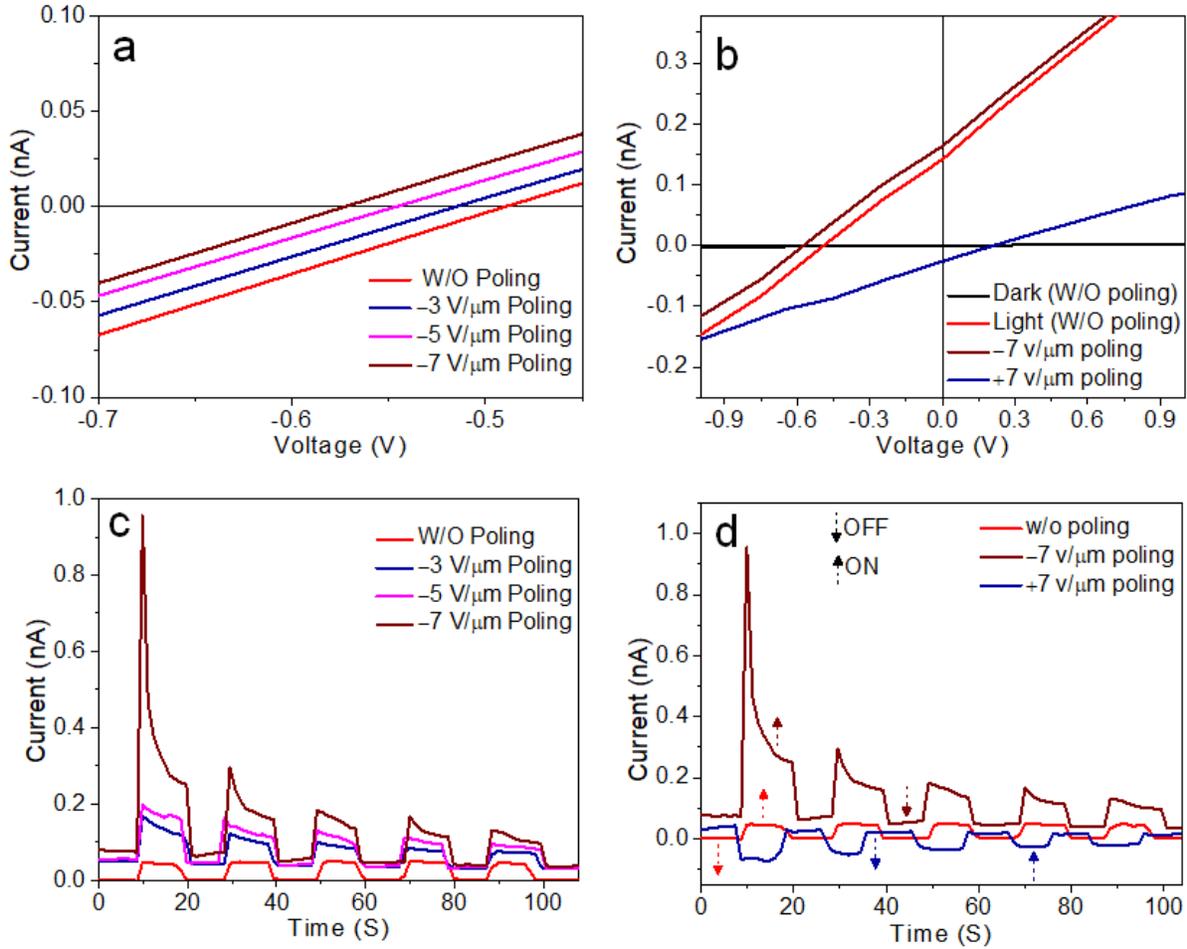

**Figure 3.** (a) Current versus voltage curves at zero-bias and after poling with increasing electric field. Light intensity was 8 mW/cm$^2$ (b) $V_{OC}$ and $I_{SC}$ switching after changing the direction of electrical poling. Light intensity was 8 mW/cm$^2$ (c) The time-dependent light on-off response after poling with increasing electric field. Light intensity was 5 mW/cm$^2$ (d) Transient photoresponse switching under zero-bias and after poling with revered electric field. Light intensity was 5 mW/cm$^2$.

To determine the origin of the switchable photovoltaic effect in $CsPbBr_3$ NCs, we accounted possible mechanisms which have been reported for switchable photovoltaic behavior earlier.[61] These include charge trapping in the active layer surface,[62,63] ion drift under electric field[33,64] and ferroelectricity of the photoactive layer.[65] The charge trapping mechanism in $CsPbBr_3$ NCs can be excluded because it cannot explain flipping in the direction of the photovoltage and photocurrent. The photovoltage and photocurrent obtained at the zero-bias condition overruled the ion drift mechanism. The changed photovoltage with respect to the electrode spacing (30 μm, 50 μm, 60 μm and 80 μm) in the lateral structure devices supports existence of ferroelectric photovoltaic effect (Figure S12).

A steady photovoltage or photocurrent is generated along the polarization direction of a ferroelectric material upon illumination. Hence the $CsPbBr_3$ NCs not only detects light to act as a photodetector but also results in a photovoltage for the photovoltaic effect in absence of any external field thus making it a self-powered device. The charge carriers are separated through the local electric fields originating from the electric polarization of ferroelectric NCs with inherent non-centrosymmetry resulting in ferroelectric photovoltaic effect.[66] Such light sensing performances of $CsPbBr_3$ NCs are characterized from the figure-of-merits such as responsivity (R) and detectivity (D*) at zero-bias conditions (Figure 4a). The R and D* are measured to be 0.14 mA/W and $0.42 \times 10^{12}$ Jones at $0.5$ mW/cm$^2$ respectively. We have also tested R and D* at 10 V bias condition (Figure S13). The R and D* were estimated to be 3.24 mA/W and $2.41 \times 10^{12}$ Jones respectively. The D* is comparable to commercially available Si based photodetectors ($4 \times 10^{12}$ Jones). Response time is another important figure-of-merits, which describes the ability of a device to switch in response to high-speed optical signals. The transient characteristics of a representative single cycle illustrate the saturation current in the on-state and constant current in the off-state (Figure 4b). The rise time and decay time extracted from the rise and decay edges of the light on-off cycle indicate a rapid photoresponse characteristics at zero-bias. The rising and falling edges of the CsPbBr$_3$ NCs photoresponse display a linear relationship with two time constants for the light on-off processes suggesting surface recombination of photogenerated carriers and their capture in surface states were involved in the photocurrent.[47] The rise and decay times are calculated to be 44.3ms and 47 ms, respectively indicating a fast photoresponse. Flexibility and environment fatigue durability are crucial for practical applications of photodetectors.[67] The flexibility of the photodetectors was monitored by drop casting the CsPbBr$_3$ NCs on a flexible PET substrate in Ag/CsPbBr$_3$ NCs/Ag device configuration. The time-dependent photoresponse shows photodetection for multiple light on-off cycles at zero-bias, which increases with the light intensity revealing the potential to fabricate photodetector on flexible PET substrate. Furthermore, the photoresponse was measured by flexing the device for multiple bending cycles. We have conducted the durability fatigue tests by recording the photocurrent and dark current of the device over the course of 1000 bending cycles (Figure 4d). The photoresponse is effectively unchanged, indicating excellent robustness, in addition to flexibility. The photostability of the devices was examined by performing long term photoresponse measurements with arbitrary light on-off cycles (Figure 4e). The device exhibits reproducible and quick response to periodic intermittent light after exposing to white light over 15 hr. It demonstrates a steady current under continuous light illumination demonstrating robust photostability. The device also maintained a constant current for prolong time upon continuous illumination in absence of any applied external electric field, which makes the device behaving like a current source at zero-bias conditions (Figure 4f).

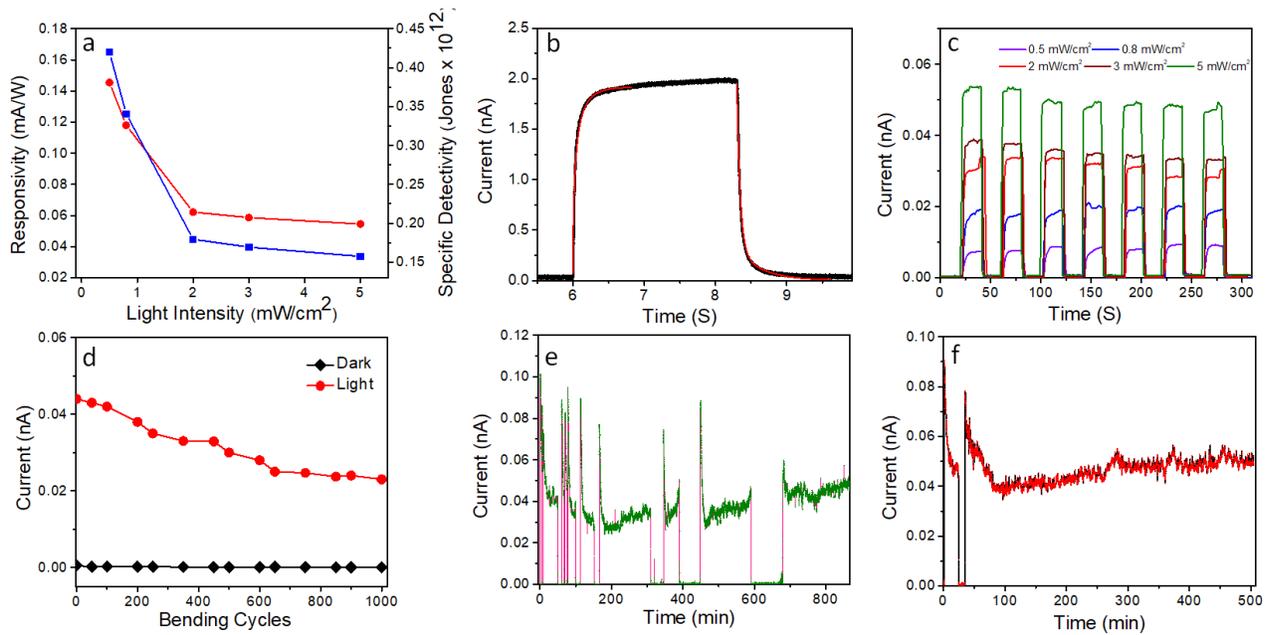

**Figure 4.** (a) Responsivity (red) and detectivity (blue) as a function of light intensity at zero-bias conditions. (b) Rise time and fall time of a single light on-off cycle. Black dots and red curves represent experimental data and exponential fits. (c) Transient photoresponse for the CsPbBr$_3$ NCs on flexible PET substrates at zero-bias condition. (d) Photocurrent and dark current measured for 1000 bending cycles. (e) Time-dependent random light on-off photoresponse at zero-bias conditions showing long term photoresponse over 15 h. (f) Prolonged photoresponse under continuous light illumination with 5 mW/cm$^2$ intensity.

## 3. Conclusion

In conclusion, we have shown a ferroelectricity-driven photovoltaic effect in all-inorganic perovskite $CsPbBr_3$ NCs. We convincingly probed ferroelectricity using the local and bulk measurements of polarization in the $CsPbBr_3$ NCs. High-temperature ferroelectric $CsPbBr_3$ NCs with a $T_c$ of 465 K (75 K above that of BTO) makes it attractive for photovoltaics applications without using an external electric field. The photovoltaic activity is ascribed to the independent separation and transport of photon-generated carriers associated with the intrinsic ferroelectricity in $CsPbBr_3$ NCs. The lateral structure devices with symmetric electrodes are particularly interesting since it eliminates the need for asymmetric electrodes or need of external electric field for charge separation. The devices demonstrated large on-off ratio of the photocurrent, flexibility, robust fatigue and prolonged photoresponse at zero-bias conditions. The photovoltaic performance was further enhanced by poling the $CsPbBr_3$ NCs with an electric field in a controllable manner.

Furthermore, the field-switchable photovoltaic was achieved by changing the direction of poling providing additional flexibility to the devices. These findings reveal that the strategy of integrating ferroelectricity into all-inorganic perovskite NCs would serve as a promising approach for fabrication of high-performance photovoltaic devices.

## Supporting Information

Supporting Information is available from the Wiley Online Library or from the author.

## Acknowledgements

The authors thank SERB-STAR grant #STR/2020/000053 India for financial support. The authors gratefully acknowledge Prof. Devajyoti Mukherjee, SPS of IACS for the preliminary macroscopic dielectric and ferroelectric measurements. A. G. acknowledges DST INSPIRE for the fellowship. The authors acknowledge the technical research centre (TRC) and Prof. Mintu Mondal of SPS of IACS for the instrumental support.

## Conflict of Interest

The authors declare no conflict of interest

## Keywords

$CsPbBr_3$ NCs, High $T_c$, ferroelectricity, bulk photovoltaics, switchable polarization,

**TOC**

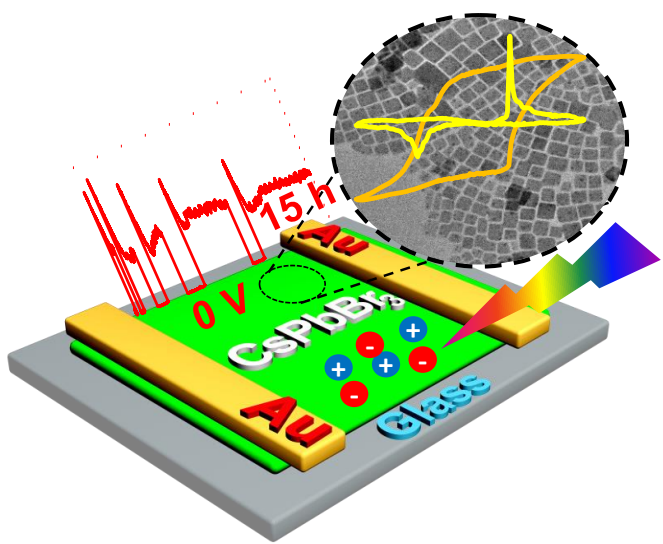

# Supporting Information

# Switchable Photovoltaic Effect in Ferroelectric CsPbBr$_3$ Nanocrystals

Anashmita Ghosh[1], Susmita Paul[1], Mrinmay Das[1], Piyush Kanti Sarkar[1], Pooja Bhardwaj[2], Goutam Sheet[2], Surajit Das[1], Anuja Datta[1] and Somobrata Acharya[1]*

[1]School of Applied & Interdisciplinary Sciences, India Association for the Cultivation of Science, Jadavpur, Kolkata-700032, India

[2]Department of Physical Sciences, Indian Institute of Science Education and Research (IISER) Mohali, Sector 81, S. A. S. Nagar, Manauli P.O. 140306, India

Corresponding Author: camsa2@iacs.res.in

## Contents

**Experimental Section**

**SEM and TEM measurements**

SEM images were acquired using a JEOL JSM-7500F field-emission scanning electron microscope. Transmission electron microscopy (TEM) was performed using UHR-FEG-TEM, JEOL JEM-2010 electron microscopy operating at 200 kV electron source. Selected area electron diffraction (SAED) and energy dispersive X-ray spectroscopy (EDS) were obtained with the same electron microscope. Powder XRD measurements were carried out using Rigaku SmartLab diffractometer using Cu Ka ($\lambda$ = 1.54 Å) as the incident radiation.

**Device fabrication**

The CsPbBr$_3$ NCs based photovoltaic devices were fabricated in metal-semiconductor-metal (MSM) lateral configuration. A solution of as-synthesized CsPbBr$_3$ NCs in hexane was prepared for the thin film preparation on different substrates. CsPbBr$_3$NCs solution was drop casted onto the precleaned substrates and then dried in vacuum. To fabricate the MSM devices, Au (50 to 100 nm) or Ag (50 to 100 nm) electrodes were deposited on the film of NCs by thermal evaporation technique using a Hind High Vacuum Co., India with a shadow mask purchased from Ossila Limited. The devices were fabricated with different channel lengths (gap between the electrodes) varying from 30 μm to 80 μm. The width of the channels was fixed at 1 mm.

**Current−voltage characteristic measurements**

The current−voltage characteristic and transient photoresponse measurements were carried using a TTPx Lakeshore probe station connected to a Keithley 2635B source-measure unit. For photovoltaic or transient characteristics measurements Optem Schott light source was used. Electrical poling was carried out using a high voltage DC power supply from Scientific Instruments. The frequency response measurements were conducted with a violet laser diode (410 nm) and oscilloscope from Scientific Instruments. The flexibility tests were performed using a home-built bending apparatus in tandem with the probe station.

**Dielectric and ferroelectric measurements**

Capacitor-type devices were prepared for the macroscopic dielectric and ferroelectric measurements. Macroscopic dielectric and ferroelectric measurements (P-E loop and PUND test) were carried out using a Radiant Technologies Precision Multiferroic tester and performed using tungsten (W) probes to contact with the fabricated metallic electrodes. Dielectric measurements were carried out in the frequency range of 2 KHz to 1 MHz at temperature range of 290 K to 673 K. Ferroelectric properties and leakage current were evaluated at different temperatures (up to 673 K).

**PFM and local switching spectroscopy**

PFM measurements were performed using a scanning probe microscope (Asylum, MFP-3D). Dual-AC resonance tracking mode of PFM was conducted using a conductive Pt/Ir-coated probe tip (NanoSensor, PPP-EFM) to image the domain structures, to measure switching spectroscopy and piezoelectric hysteresis loops. The topographic AFM images were obtained simultaneously from the contact mode of PFM scanning. For the vector PFM containing two in-plane PFM and one out-of-plane PFM imaging, the sample was rotated by 90° between the two in-planes of PFM measurements.

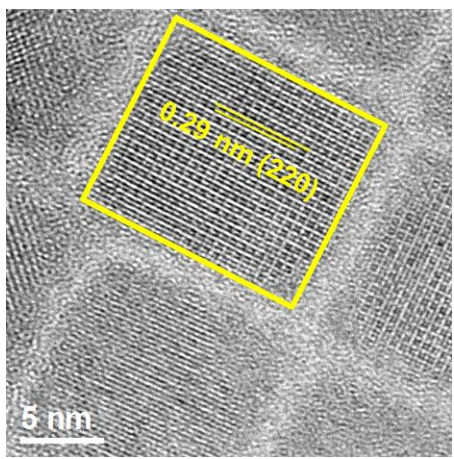

**Figure S1.** High resolution TEM (HRTEM) image of $CsPbBr_3$ NCs showing lattice planes. HRTEM image shows an interplanar spacing of ~0.29 ± 0.05 nm corresponding to the (220) planes of bulk orthorhombic crystal structure (COD # 1533062).

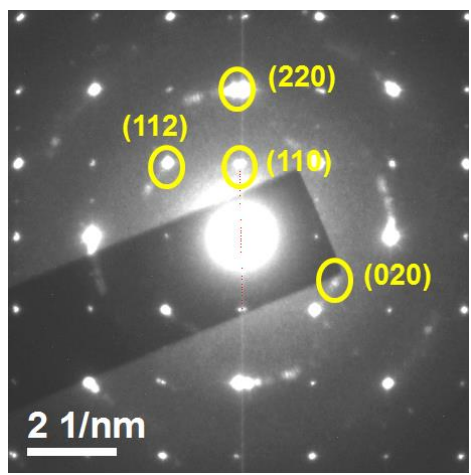

**Figure S2.** Selected area electron diffraction (SAED) pattern of the CsPbBr$_3$ NCs showing diffraction spots corresponding to the (112), (220), (110) and (020) planes of orthorhombic crystal phase.

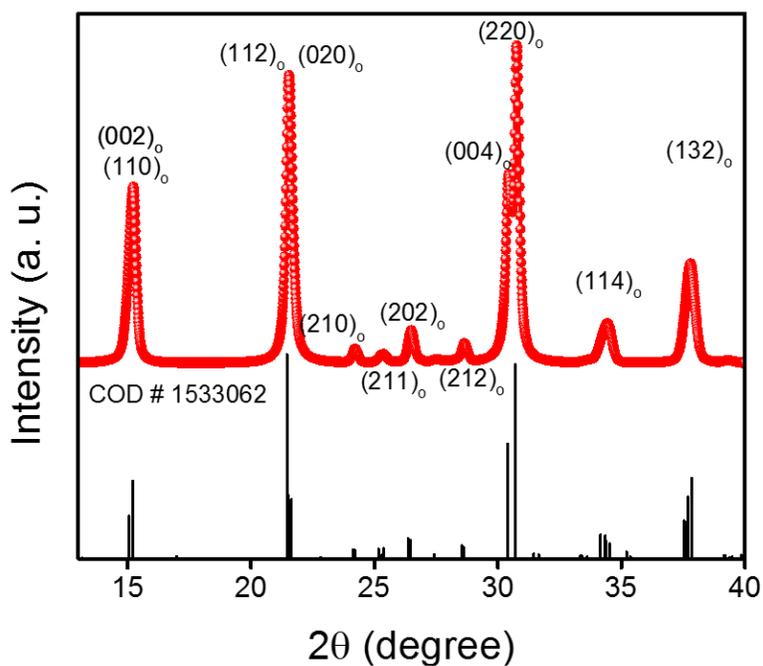

**Figure S3.** Powder X−ray diffraction (XRD) pattern of CsPbBr$_3$ NCs showing reflections corresponding to orthorhombic phase (COD # 1533062; space group: Pbnm (62); a = 8.2070 Å, b = 8.2550 Å, c = 11.7590 Å).

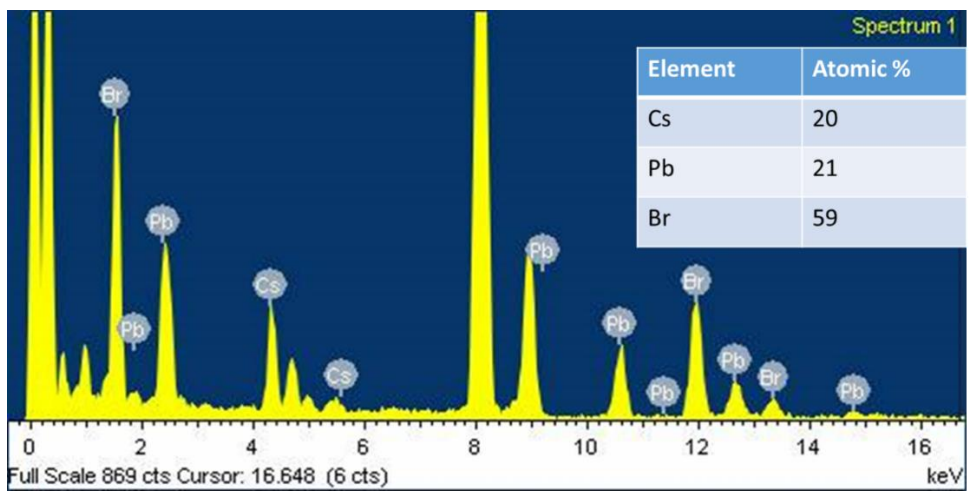

**Figure S4.** Energy dispersive X–ray spectroscopy (EDS) measurements in TEM revealing atomic ratio of Cs:Pb:Br ~ 1:1:3, which confirms the $CsPbBr_3$ composition.

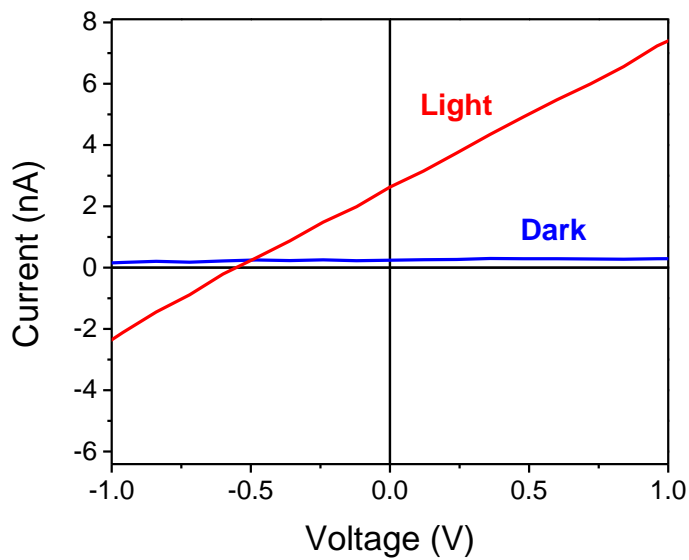

**Figure S5.** Current versus voltage curves measured under light (red) and dark (blue) at zero-bias condition for the device fabricated with Au/Ti/ $CsPbBr_3$/Ti/Au structure.

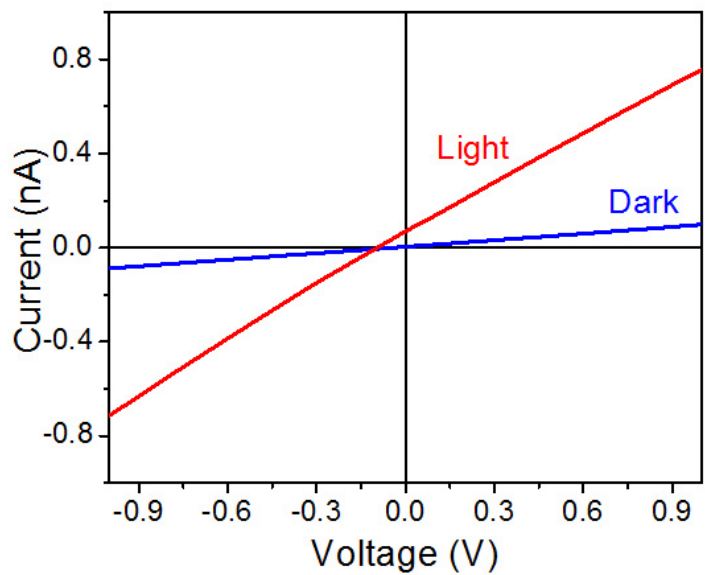

**Figure S6.** Current versus voltage curves measured under light (red) and dark (blue) at zero-bias conditions for the device fabricated on PET substrate with Ag/ CsPbBr3/Ag structure.

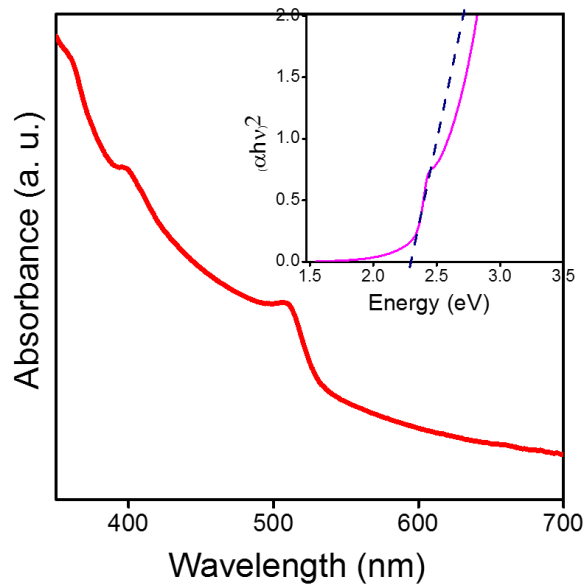

**Figure S7**. UV-vis absorption spectrum of CsPbBr3 NCs in hexane solution. The inset shows the Tauc plot. Absorption spectra were measured using Varian Carry 5000 UV−vis-NIR spectrophotometer.

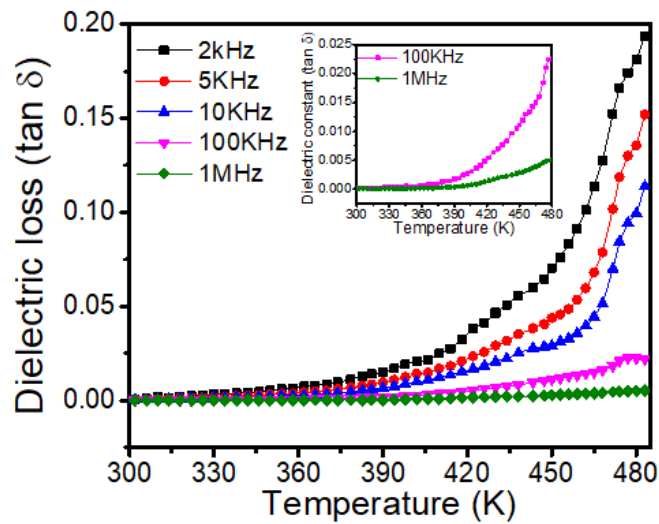

**Figure S8**. Dielectric loss versus temperature curves of CsPbBr$_3$ NCs measured at different frequencies.

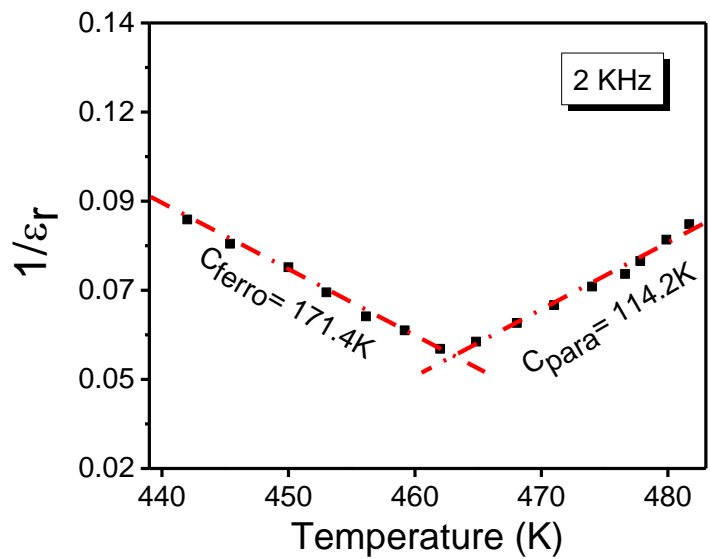

**Figure S9**. Plot of Curie–Weiss law showing the reciprocal of $\varepsilon_r$ as a function of temperature is linear.

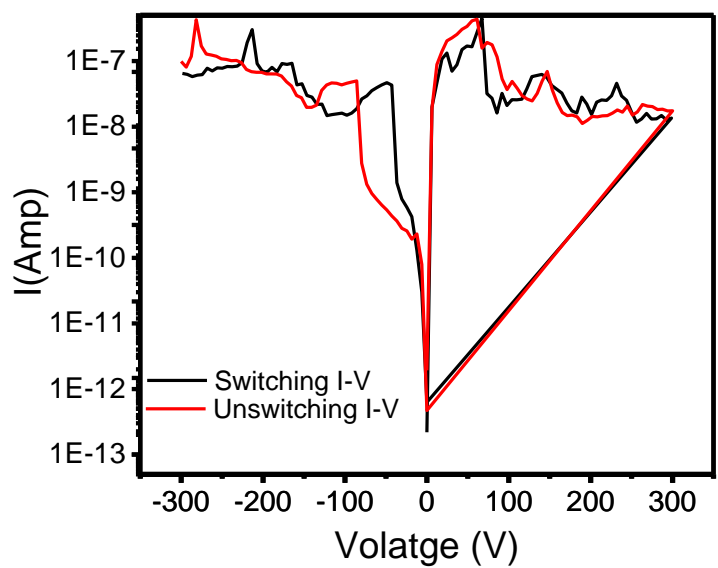

**Figure S10.** Leakage switching current versus voltage curves showing a low leakage current of the ferroelectric CsPbBr$_3$ NCs indicating the efficacy of the thin-film capacitor structures.

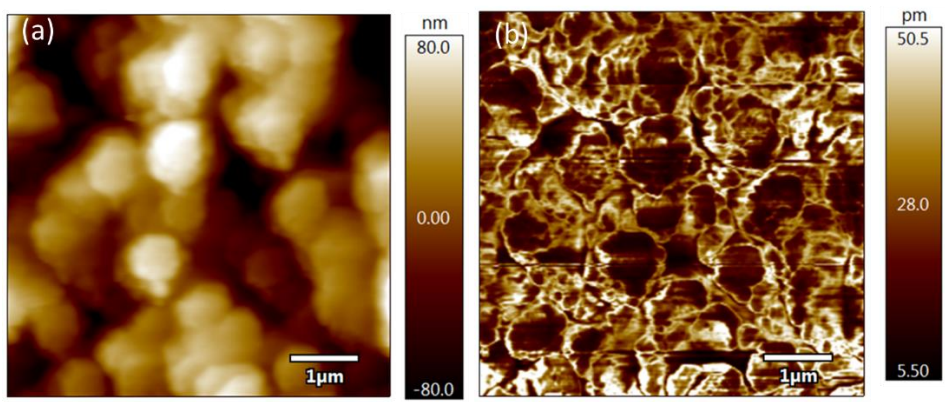

**Figure S11.** PFM (a) topography image (b) Amplitude image of CsPbBr$_3$ NCs.

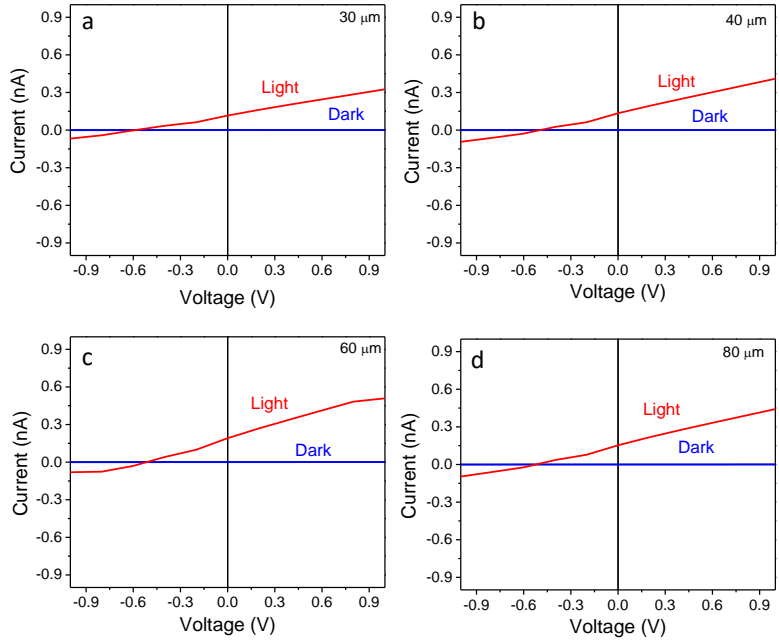

**Figure S12.** Current versus voltage curves measured under light (red) and dark (blue) with lateral electrode spacings' of (a) 30 μm (b) 40 μm (c) 60 μm and (d) 80 μm respectively. The intensity of the illuminated light was 8mW/cm$^2$.

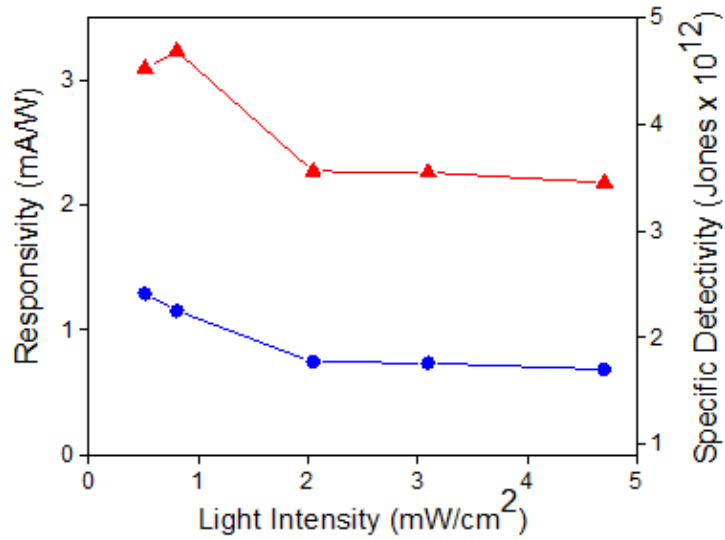

**Figure S13.** Responsivity (red) and detectivity (blue) as a function of light intensity at 10 V bias.